\input harvmac
\input epsf
\def\lf{16\pi^2}

\def\frak#1#2{{\textstyle{{#1}\over{#2}}}}
\def\frakk#1#2{{{#1}\over{#2}}}
\def\pa{\partial}
\def\Acal{{\cal A}}
\def\Bcal{{\cal B}}
\def\Ccal{{\cal C}}
\def\Dcal{{\cal D}}
\def\semi{;\hfil\break}
\def\ga{\gamma}

\def\Xtilde{\tilde X}
\def\Ytilde{\tilde Y}
\def\sy{supersymmetry}

\def\NSVZ{{\rm NSVZ}}
\def\DRED{{\rm DRED}}
\def\DREDp{{\rm DRED}'}

\def\npb{{Nucl.\ Phys.\ }{\bf B}}

\def\prd{{Phys.\ Rev.\ }{\bf D}}

\def\plb{{Phys.\ Lett.\ }{\bf B}}
\def\ijmpa{{Int.\ J.\ Mod.\ Phys.\ }{\bf A}}
\def\mtilde{\tilde m}

\def\bM{M^*}

\def\drbar{{\overline{\rm DR}}}
\def\lf{16\pi^2}
\def\llf{(16\pi^2)^2}
\def\lllf{(16\pi^2)^3}
\def \in{\leftskip = 40 pt\rightskip = 40pt}
\def \out{\leftskip = 0 pt\rightskip = 0pt}
{\nopagenumbers
\line{\hfil LTH 426}
\line{\hfil hep-ph/9805482}
\vskip .5in    
\centerline{\titlefont The connection between the
DRED and }
\vskip 10pt
\centerline{\titlefont NSVZ Renormalisation Schemes\foot{This paper is 
dedicated to the memory of Mark Samuel}}
\vskip 1in
\centerline{\bf I.~Jack, D.R.T.~Jones and A.~Pickering}
\medskip
\centerline{\it Dept. of Mathematical Sciences,
University of Liverpool, Liverpool L69 3BX, UK}
\vskip .3in

We explore the relationship between the DRED and NSVZ schemes. Using 
certain exact results for the soft scalar mass $\beta$-function, 
we derive the transformation of $\alpha^{\NSVZ}$ to $\alpha^{\DRED}$ 
through terms of order $\alpha^4$. We thus incidentally determine 
$\beta_{\alpha}^{\DRED}$ through four loops, and we compare our result to a 
previous Pad\'e Approximant prediction.
\Date{ May 1998}}

In a recent series of papers we have explored the scheme--dependence 
associated with the renormalisation 
of the coupling constants and mass-parameters of a softly-broken
supersymmetric theory. There are two schemes of particular interest, 
which we term the DRED scheme and the NSVZ scheme. 
The DRED scheme is defined by the procedure of minimal (or modified minimal) 
subtraction associated 
with regularisation by dimensional reduction; also known as DR (or $\drbar$).
(The distinction between DR and $\drbar$ is immaterial for our present 
purposes.) The NSVZ scheme is one such that the 
NSVZ formula~\ref\NSVZb{V.~Novikov, M.~Shifman, A.~Vainstein and V.~Zakharov, 
\npb 229 (1983) 381\semi
V.~Novikov, M.~Shifman, A.~Vainstein and V.~Zakharov, \plb166 (1986) 329\semi
M.~Shifman and A.~Vainstein, \npb 277 (1986) 456} 
for the gauge $\beta$-function $\beta_{\alpha}$ holds; we will define this 
scheme in more detail later. 
The NSVZ formula relates $\beta_{\alpha}$
to the the anomalous dimension
matrix $\ga$ of
the chiral superfields as follows: 

\eqn\An{
\beta_{\alpha}^{\NSVZ}=2{\alpha^2\over{16\pi^2}}
\left[{Q-2r^{-1}\tr[\gamma C(R)]\over{
1-2\alpha C(G)(\lf)^{-1}}}\right] ,}
where $\alpha = g^2$ and $Q = T(R) - 3C(G)$. 
$C(R)$ and $C(G)$ are the quadratic matter and adjoint Casimirs respectively; 
$T(R) = r^{-1}d_R\tr[C(R)]$, where $d_R$ is the dimension of the matter 
representation, and  
$r$ is the number of generators of the gauge group. 

The NSVZ scheme is related to the ``holomorphic'' scheme (wherein 
the one--loop $\beta_{\alpha}$ is exact) by the 
transformation

\eqn\hol{
{1\over{\alpha^{{\rm H}}}}={1\over{\alpha^{\NSVZ}}}+\frakk{2}{\lf}
C(G)\ln\alpha^{{\NSVZ}}
-\frakk{4}{\lf}r^{-1}\tr [ZC(R)],} where $\mu\frakk{dZ}{d\mu} = \ga$.

No relation of the form of Eq.~\An\ exists in DRED; on the other hand, 
DRED is a well-defined procedure for the calculation of radiative 
corrections. Thus if we wish to perform a calculation which involves 

(1) Running couplings and masses from $10^{16}$GeV (say) to 
$M_Z$ and then

(2) Calculating radiative corrections to physical masses and 
processes,

{\parindent = 0pt we might well consider using NSVZ (or the holomorphic scheme) 
in the 
former procedure and DRED in the latter. In fact, the NSVZ scheme has been used 
for running the dimensionless couplings in Refs.~\ref\shif{M.~Shifman,
\ijmpa11 (1996) 5761}, \ref\ross{G.~Amelino-Camelia, D.~Ghilencea and
G.G.~Ross, hep-ph/9804437} (see also Ref.~\ref\mur{N.~Arkani-Hamed 
and H.~Murayama, \prd57 (1998) 6638}). For this reason it is useful to 
know as precisely as possible the connection between the schemes.}

In  Refs.~\ref\jjna{I.~Jack,
D.R.T.~Jones and  C.G.~North, \plb 386 (1996) 138}, \ref\jjn{I.~Jack,
D.R.T.~Jones and  C.G.~North, \npb486 (1997) 479} we constructed
perturbatively a redefinition  $\alpha^{\NSVZ}\to\alpha^{\DRED}$ 
by comparing $\beta_{\alpha}$  in the two schemes. The result was 
\eqn\redefa{
\alpha^{\DRED}= \alpha^{\NSVZ}+\sum_{L=1}^{\infty}
\delta^{(L)}(\alpha^{\NSVZ},Y,Y^*),}
where $\delta^{(1)}=0$, 
\eqn\tlfbb{
\llf\delta^{(2)}=\alpha^2
\left[r^{-1}\tr\left[PC(R)\right]-\alpha QC(G)\right],}
and
\eqn\redefb{
\delta^{(3)}=\rho_1\Delta_1+ \rho_2\Delta_2+ \rho_3\Delta_3.}
Here
\eqna\fourb$$\eqalignno{ \lllf\Delta_1&=
\alpha^3C(G)\left[r^{-1}\tr[PC(R)] - \alpha QC(G)\right]&\fourb a\cr
\lllf\Delta_2&= r^{-1}\tr\left[\alpha^2S_4C(R) -2 \alpha^4QC(R)^2 + 2   
\alpha^3PC(R)^2\right]  &\fourb b\cr \lllf\Delta_3&=
\alpha^2r^{-1}\tr[P^2C(R)] - \alpha^4 Q^2C(G), &\fourb c\cr}$$
and in Ref.~\jjn\ we showed that $\rho_2 = -\frakk{4}{3}$ and 
$\rho_3 = \frakk{1}{3}$.
$P^i{}_j$ and $S_4 ^i{}_j$ are defined as follows:

\eqn\eqf{\eqalign{P^i{}_j &=\frak{1}{2}Y^{ikl}Y_{jkl}-2\alpha C(R)^i{}_j,\cr 
S_4 ^i{}_j &= Y^{imn}P^p{}_m Y_{jpn},\cr}}
where we have written the superpotential as
\eqn\eqg{W(\Phi) = \frak{1}{6}Y^{ijk}\Phi_i\Phi_j\Phi_k
+\frak{1}{2}\mu^{ij}\Phi_i\Phi_j.}
As usual we raise and lower indices by complex conjugation, e.g. 
$Y_{ijk}=(Y^{ijk})^*$.
In principle the undetermined coefficient $\rho_1$  could
be found by the same method as employed in Ref.~\jjn\ 
to find $\rho_2$ and $\rho_3$; that is, by 
calculating a relevant contribution to 
$\beta_{\alpha}^{(4)\DRED}$. This would be very tedious, however\foot{In
Ref.~\ref\jjs{I.~Jack, D.R.T.~Jones and M.A.~Samuel, \plb 407 (1997)
143} a method based on Pad\'e approximants was used to suggest that
$\rho_1\approx 4.9$}. In this paper we  show that our recent work on the
soft \sy-breaking $\beta$-functions leads to a determination of $\rho_1$
based on a remarkably simple three-loop calculation.

We take the soft breaking Lagrangian $L_{SB}$ as follows:
\eqn\Aba{
L_{SB}(\phi,\lambda) = \left[{1\over6}h^{ijk}\phi_i\phi_j\phi_k
+{1\over2}b^{ij}\phi_i\phi_j+{1\over2}M\lambda\lambda
+{\rm h.c.}\right]-(m^2)^i{}_j\phi^j\phi_i.}
Here $M$ is the gaugino mass, and $\phi^i = \phi^*_i$.
In Ref.~\ref\jjp{I.~Jack, D.R.T.~Jones and A.~Pickering, hep-ph/9712542}\ 
we showed that the soft scalar mass $\beta$-function 
is given by the following expression:
\eqn\Ajy{
(\beta_{m^2})^i{}_j=\left[ \Delta + 
\Xtilde(\alpha, Y, Y^*, h, h^*, m, M)\frakk{\pa}{\pa \alpha}\right]
\gamma^i{}_j,}
where
\eqn\Ajz{
\Delta = 2{\cal O}{\cal O}^* +2M\bM \alpha{\pa
\over{\pa \alpha}} +\Ytilde_{lmn}{\pa\over{\pa Y_{lmn}}}
+\Ytilde^{lmn}{\pa\over{\pa Y^{lmn}}},}
\eqn\Ajb{
{\cal O}=\left(M\alpha{\pa\over{\pa \alpha}}-h^{lmn}{\pa
\over{\pa Y^{lmn}}}\right),}
and
\eqn\Ajd{
\Ytilde^{ijk}=(m^2)^i{}_lY^{ljk}+(m^2)^j{}_lY^{ilk}+(m^2)^k{}_lY^{ijl}.}
The function $\Xtilde$ was introduced  in Ref.~\jjp; it does not appear in a
naive application of the spurion  formalism~\ref\avd{L.V.~Avdeev,
D.I.~Kazakov  and I.N.~Kondrashuk, \npb510 (1998) 289},  because (when
using DRED) this fails to allow for the fact  that the
$\epsilon$-scalars associated with DRED acquire a mass through radiative
corrections~\ref\jj{I.~Jack and D.R.T.~Jones, \plb 333 (1994) 372}. (For
further discussion, see  Refs.~\ref\jjg{I.~Jack and D.R.T.~Jones, \plb 415
(1997) 383}, \ref\aglr{N.~Arkani-Hamed, G.F.~Giudice, M.A.~Luty and
R.~Rattazzi, hep-ph/9803290}.) 
Indeed, in DRED, $\beta_{m^2}$ will actually depend on the
$\epsilon$-scalar mass. It is, however, possible to  define a scheme,
$\DREDp$, related to DRED, such that $\beta_{m^2}$ is independent of the
$\epsilon$-scalar mass \ref\jjmvy{I.~Jack, D.R.T.~Jones, S.P.~Martin,
M.T.~Vaughn and Y.~Yamada, \prd50 (1994) R5481}.

In Ref.~\ref\jjpb{I.~Jack, D.R.T.~Jones and A.~Pickering, hep-ph/9803405} we
claimed that Eq.~\Ajy\ holds in both $\DREDp$  and the NSVZ
scheme, the two schemes being related by  the transformation
Eq.~\redefa\ and an associated transformation on  the gaugino mass $M$,
given by 
\eqn\Ar{
\alpha M
=\alpha^{\prime}M^{\prime}{{\pa \alpha(\alpha',Y,Y^*)}
\over{\pa \alpha'}}
-h^{ijk}{{\pa \alpha(\alpha',Y,Y^*)}\over{\pa Y^{ijk}}}.}
(These two transformations define precisely what we mean by the NSVZ scheme.)
We also argued that in the NSVZ scheme we have simply
\eqn\exX{
\Xtilde^{\NSVZ}=-4{\alpha^2\over{16\pi^2}}
{S\over{\left[1-2\alpha C(G)(16\pi^2)^{-1}\right]}}}
where 
\eqn\Awc{
S =  r^{-1}\tr [m^2C(R)] -MM^* C(G),}
whereas in the $\DREDp$ scheme, $\Xtilde$ is related to the $\beta$-function 
for the $\epsilon$-scalar mass, $\mtilde$. Writing 
\eqn\betatm{
\beta_{\mtilde^2}=N_1+N_2\mtilde^2,}
where $N_1(\alpha,Y,Y^*,h,h^*,m,M)$ 
does not depend on $\mtilde$, we have 
\eqn\Xdred{
\Xtilde^{\DREDp} =-\sum_{L=1}^{\infty}{\alpha\over L}N_1^{(L)}}
where $N_1^{(L)}$ is the $L$-loop contribution to $N_1$. The distinction
between DRED and $\DREDp$ has no influence on the calculation  of $N_1$;
the $\DRED\to\DREDp$ redefinition only changes $N_2$. 

Clearly, given Eqs.~\Xdred\ and \exX, we can determine the relation between
$\alpha^{\DRED}$ and $\alpha^{\NSVZ}$ by calculating $N_1$ if we know 
how $\Xtilde$ transforms under a scheme redefinition. We showed in Ref.~\jjpb\
that under 
a transformation $\alpha\rightarrow\alpha'$, with an associated transformation
of $M$ given by Eq.~\Ar, the transformed $\Xtilde'$ is related to $\Xtilde$ by  
\eqn\delXa{\eqalign{
\Xtilde= &\Xtilde'{\pa \alpha\over{\pa \alpha'}}+
2M'M^{\prime*}\left[\alpha^{\prime2}{\pa^2\alpha\over{\pa
\alpha^{\prime2}}}
+2\alpha'{\pa\alpha\over{\pa\alpha'}}-2\frakk{\alpha^{\prime2}}{\alpha}
\left(\frakk{\pa\alpha}{\pa\alpha'}\right)^2\right]\cr
&-\left[2M'\alpha'h_{ijk}\left({\pa^2\alpha\over{\pa Y_{ijk}\pa\alpha'}}
-\frakk{2}{\alpha}\frakk{\pa\alpha}{\pa Y_{ijk}}\frakk{\pa\alpha}{\pa\alpha'}
\right)-\Ytilde^{ijk}{\pa\alpha\over{\pa Y^{ijk}}}+\hbox{c.c.}\right] \cr
&+2h^{ijk}h_{lmn}\left[{\pa^2\alpha\over{\pa Y^{ijk}\pa Y_{lmn}}}
-\frakk{2}{\alpha}\frakk{\pa\alpha}{\pa Y^{ijk}}
\frakk{\pa\alpha}{\pa Y_{lmn}}\right]
,\cr}}
and we also showed that Eq.~\delXa\ is consistent with Eq.~\tlfbb. Moreover, 
we calculated the contributions of all tensor structures  of the general
form $\alpha m^2Y^2Y^{*2}C(R)$  
to $N_1$, which enabled us to test Eq.~\delXa\ against
Eq.~\redefb;    this calculation did not involve tensor structures 
associated with $\rho_1$. Our confidence bolstered by this, we now proceed to
determine $\rho_1$ by calculating the  $N_1$ contributions from some
tensor structures that {\it are\/}  sensitive to $\rho_1$.  This involves
a three-loop calculation  in the broken theory; far simpler than the
{\it four}-loop calculation  (albeit in the unbroken theory) required to
determine $\rho_1$ from  $\beta^{\DRED}_{\alpha}$. 

If we take the primed scheme  in Eq.~\delXa\ to be NSVZ and the unprimed
scheme to be $\DREDp$, then  using Eqs.~\redefa-\redefb\ we find that 
in $\DREDp$

\eqn\alphone{\eqalign{\lllf\Xtilde^{(3)} &= 
r^{-1}\alpha^3C(G)\Bigl[ \rho_1\tr\left[WC(R)\right] 
+(4\rho_1 -8)MM^*\tr\left[PC(R)\right]\cr
&-12r(\rho_1 -1)\alpha MM^*QC(G)+4\alpha Q\tr\left[m^2C(R)\right]\cr
&-(\rho_1 -2)\left\{M^*\tr\left[HC(R)\right] + \hbox{c.c.}\right\}\Bigr]
+ 8\alpha^4\tr\left[C(R)^2\right]S\cr
&+2\rho_2 r^{-1}\alpha^3\tr\left[\left\{W-M^*H-MH^*
+4MM^*P-12\alpha MM^*Q\right\}C(R)^2\right]\cr
&+\rho_2 r^{-1}\alpha^2\Bigl[Y^{ijk}W^l{}_k C(R)^m{}_j Y_{ilm}
+ 2h^{ijk}P^l{}_kC(R)^m{}_j h_{ilm}\cr
&+\left(Y^{ijk}H^l{}_k C(R)^m{}_j h_{ilm} +
Y^{ijk}P^l{}_k C(R)^m{}_j \Ytilde_{ilm} + \hbox{c.c.}
\right) \Bigr]-16\alpha^4SC(G)^2\cr
&+\rho_3 r^{-1}\alpha^2\left[ \tr\left[HH^* C(R)\right] - 12r\alpha^2 
MM^* Q^2 C(G) 
+2\tr\left[WPC(R)\right]\right] \cr}}

where
\eqn\Aaa{
W^j{}_i= \frak{1}{2}\left[ Y^{jkl}\Ytilde_{ikl} 
+ \Ytilde^{jkl}Y_{ikl}\right] +h_{ipq}h^{jpq}-8\alpha MM^*C(R)^j{}_i,}
and
\eqn\Aab{
H^i{}_j=h^{ikl}Y_{jkl}+4\alpha MC(R)^i{}_j.}
(The combination $H^i{}_j$ was formerly known~\jj\ as $X^i{}_j$.) 
We have not substituted for $\rho_2$ and $\rho_3$ in Eq.~\alphone\ above 
so that the contributions emanating from the first term on the RHS 
of Eq.~\delXa\ are easier to identify. 
As explained above, in Ref.~\jjpb\ we explicitly calculated the 
three-loop Feynman diagrams corresponding to contributions  to
$\beta_{\mtilde^2}$ of the form $\alpha m^2Y^2Y^{*2}C(R)$.  Here we report on
analogous calculations involving the following tensor structures:  
\eqn\struct{\eqalign{
T_1&= \alpha^2h^*MYC(R)C(G) + \hbox{c.c},\cr
T_2&= \alpha^2hh^*C(R)C(G),\cr
T_3&=\alpha^2MM^*YY^*C(R)C(G).}}
These all have coefficients that depend on $\rho_1$. 

The Feynman graphs giving $T_1$-type contributions to $N_1$ are  shown
in Fig.~1. 
\bigskip
\epsfysize= 1.0in
\centerline{\epsfbox{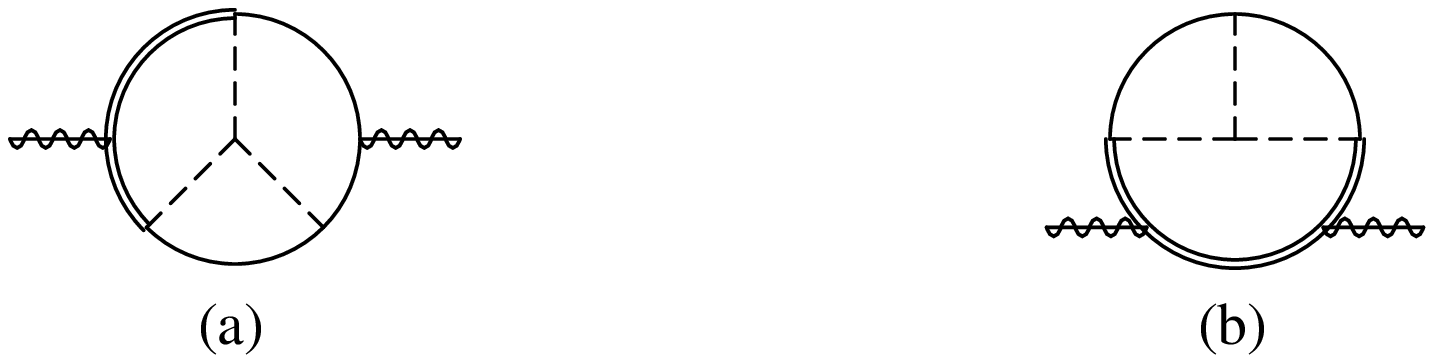}}
\in
{\it \noindent Fig.~1:
Feynman diagrams for $T_1=\alpha^2h^*MYC(R)C(G) + \hbox{c.c}$. Double lines, 
plain lines and broken lines represent the gaugino, chiral fermion and chiral
scalar propagators respectively. The external lines are $\epsilon$-scalars.}
\bigskip
\out
The calculation is quite straightforward,  especially when
one realises that Fig.~1(a) does not  contribute, because the two
possible places for the $M$-insertion  give opposite signs and cancel.
Thus we are reduced to evaluating the  simple pole in $\epsilon=4-d$ from a
single  graph, and we obtain:

\eqn\alphtwo{2 - \rho_1 =  -\frakk{10}{3}}
whence
\eqn\alphthree{\rho_1 = \frakk{16}{3}.}

As a check, we have also calculated the $T_2$ and $T_3$ contributions. Here 
we have more Feynman diagrams, including ones with vector fields; 
care must be taken with subtractions. In both cases we also obtain 
Eq.~\alphthree.

The determination of $\rho_1$ completes the   NSVZ/DRED connection through
terms of $O(\alpha^4 )$.  In terms of  physics at $M_Z$ this facilitates
a very  accurate transformation between the two schemes in the MSSM. 
It also completes the determination of $\beta_{\alpha}^{(4)\DRED}$; see 
Eq.~(3.18) of Ref.~\jjn\ (note that $\rho_1 = 2\alpha_1$). For the 
special case of SQCD 
we have the following results:
\eqn\Pe{
\beta_{\alpha}^{\DRED} = 
2\alpha\sum_{n=1}^{\infty}\beta_n\left(\frakk{\alpha}{\lf}\right)^{n},}
where
\eqna\fourb$$\eqalignno{
\beta_1 &= N_f - 3N_c, &\fourb a\cr
\beta_2 &= \left[4N_c-{2\over {N_c}}\right]N_f
-6N_c^2, &\fourb b\cr
\beta_3 &= \left[{3\over{N_c}}-4N_c\right]N_f^2
+\left[21N_c^2-{2\over{N_c^2}}-9\right]N_f
-21N_c^3, &\fourb c\cr
\beta_4 &= \Acal + \Bcal N_f + \Ccal N_f^2 +\Dcal N_f^3. &\fourb d\cr}$$ 
Here $N_c$ is the number of colours, and
\eqn\loopa{\eqalign{
\Acal &= -102N_c^4,\cr
\Bcal  &=  132N_c^3- 66N_c-\frakk{8}{N_c}
-\frakk{4}{N_c^3},\cr
\Ccal  &= -\left[42 + 12\zeta(3)\right]N_c^2
+44+\frakk{36\zeta(3)-20}{3N_c^2},\cr
\Dcal  &= -\frakk{2}{3N_c}.\cr}}
In the case $N_f = 0$ it is interesting to compare the above DRED results 
with the exact NSVZ formula,
\eqn\An{
\beta_{\alpha}^{\NSVZ}=\frakk{-6N_c\alpha^2}{16\pi^2
\left[1-2\alpha N_c(\lf)^{-1}\right]}.}
In both cases the $\beta$-function coefficients have the same sign through 
four loops. In the NSVZ case, the series manifestly has a 
finite radius of convergence; it is not clear whether or not this is true 
in the DRED case.

Our result for $\rho_1$ represents what we at  least regard as  striking
confirmation of the Asymptotic Pad\'e Approximant  prediction~\jjs\
$\rho_1/2 = 12/5$ or $5/2$.  The error of $6.25 - 10\%$  is remarkably
small and provides further evidence for the  precocious convergence of
APAPs when applied to calculations  of $\beta$--functions in non-abelian
theories\foot{With hindsight (always reliable) we might have arrived at 
the correct result for $\rho_1$  from the clue that $\rho_2$ and
$\rho_3$  are both fractions with  $3$ in the denominator}.     

Finally we have confirmation of our exact result for $\Xtilde^{\NSVZ}$, 
Eq.~\exX. We will explore the effect of this on the running analysis 
within the NSVZ scheme elsewhere. 

\bigskip\centerline{{\bf Acknowledgements}}\nobreak

The main result of this paper is the determination of $\rho_1$;  a
prediction of this~\jjs\ was made by two of us (IJ and TJ) in  collaboration
with Mark Samuel, using his  Asymptotic Pad\'e Approximation methods
\ref\eks{J.~Ellis, M.~Karliner and M.A.~Samuel, \plb400 (1997) 176}.
Mark died on November 6 1997; we  both miss his friendship and
enthusiasm. 

TJ thanks the physicists of SLAC and UC San Diego for  hospitality and
financial support while part of this work was done.

AP was supported by a PPARC Research Grant.

\listrefs
\bye